\documentclass{aa} 
\usepackage{graphicx, psfrag}
\usepackage{txfonts}
\usepackage{color}

\renewcommand{\d}{\mathrm{d}}
\newcommand{\sers}{S{\' e}rsic }
\newcommand{\equ}[1]{Eq.~\ref{eq:#1}}
\newcommand{\fig}[1]{Fig.~\ref{fig:#1}}
\newcommand{\tab}[1]{Tab.~\ref{tab:#1}}
\newcommand{\sect}[1]{Sect.~\ref{sec:#1}}

\newcommand{\atlas}{ATLAS$^\mathrm{3D}\,$}
\usepackage{xcolor}

\begin{document}

\title{MOND implications for spectral line profiles of shell galaxies: shell formation history and mass-velocity scaling relations}
% \subtitle{ }
\titlerunning{MOND implications for spectral line profiles of shell galaxies}
\author{M. B\'{i}lek\inst{1}\fnmsep\inst{2}
\and
B. Jungwiert\inst{1}
\and
I. Ebrov\'{a}\inst{1}
\and
K. Barto\v{s}kov\'{a}\inst{1}\fnmsep\inst{3}}
%\and
%I. Ebrov\'{a}\inst{1}\fnmsep\inst{4}

\institute{Astronomical Institute, Academy of Sciences of the Czech Republic, Bo\v{c}n\'{i} II 1401/1a, CZ-141\,00 Prague, Czech Republic\\
\email{bilek@asu.cas.cz}
\and
Faculty of Mathematics and Physics, Charles University in Prague, Ke~Karlovu 3, CZ-121 16 Prague, Czech Republic
%\and 
%Department of Theoretical Physics and Astrophysics, Faculty of Science, Masaryk University, Kotl\'a\v{r}sk\'a 2, CZ-611\,37 Brno, Czech Republic
\and
Department of Theoretical Physics and Astrophysics, Faculty of Science, Masaryk University, Kotl\'a\v{r}sk\'a 2, CZ-611\,37 Brno, Czech Republic
}

\date{Received ...; accepted ...}

% 5 {} token are mandatory
%%%%%%%%%%%%%%%%%%%%%%%%%%%%%%%%%%%%%%%%%%%%%%%%%%%%%%%%%%%%%%%%% 
\abstract
%%%%%%%%%%%%%%%%%%%%%%%%%%%%%%%%%%%%%%%%%%%%%%%%%%%%%%%%%%%
%%%%%%%%%%%%%%%%%%%%%%%%%%%%%%%%%%%%%%%%%%%%%%%%%%%%%%%%%%%
% context heading (optional)
{Many ellipticals are surrounded by round stellar shells probably stemming from minor mergers. A~new method for constraining gravitational potential in elliptical galaxies has recently been suggested. It uses the spectral line profiles of these shells to measure the circular velocity at the edge of the shell and the expansion velocity of the shell itself.  MOND is an alternative to the dark matter framework aiming to solve the missing mass problem. }
%%%%%%%%%%%%%%%%%%%%%%%%%%%%%%%%%%%%%%%%%%%%%%%%%%%%%%%%%%%
%%%%%%%%%%%%%%%%%%%%%%%%%%%%%%%%%%%%%%%%%%%%%%%%%%%%%%%%%%%
% aims heading (mandatory)
{We study how the circular and expansion velocities behave in MOND for large shells.}
%%%%%%%%%%%%%%%%%%%%%%%%%%%%%%%%%%%%%%%%%%%%%%%%%%%%%%%%%%%
%%%%%%%%%%%%%%%%%%%%%%%%%%%%%%%%%%%%%%%%%%%%%%%%%%%%%%%%%%%
% methods heading (mandatory)
{ The asymptotic behavior for infinitely large shells is derived analytically. The applicability of the asymptotic results for finitely sized shells is studied numerically on a~grid of galaxies modeled with \sers spheres.}
%%%%%%%%%%%%%%%%%%%%%%%%%%%%%%%%%%%%%%%%%%%%%%%%%%%%%%%%%%%
%%%%%%%%%%%%%%%%%%%%%%%%%%%%%%%%%%%%%%%%%%%%%%%%%%%%%%%%%%%
% results heading (mandatory)
{Circular velocity settles asymptotically at a~value determined by the baryonic mass of the galaxy forming the baryonic Tully-Fisher relation known for disk galaxies. Shell expansion velocity also becomes asymptotically constant. The expansion velocities of large shells form a~multibranched analogy to the baryonic Tully-Fisher relation, together with the galactic
baryonic masses. For many -- but not all -- shell galaxies, the asymptotic values of these two types of velocities are reached under the effective radius. If MOND is assumed to work in ellipticals, then the shell spectra allow many details of the history to be  revealed about the formation of the shell system, including its age. The results pertaining to circular velocities apply to all elliptical galaxies, not only those with shells.}
% conclusions heading (optional), leave it empty if necessary 
{}


\maketitle

%%%%%%%%%%%%%%%%%%%%%%%%%%%%%%%%%%%%%%%%%%%%%%%%%%%%%%%%%%%%%%%%%55
\section{Introduction}
\label{sec:intro}
Stellar shells are arc-like features observed predominantly in elliptical galaxies \citep{MC83, tal09,ramos11, atlas3d2,  atkinson13}. They are believed to be mostly remnants of minor galactic mergers \citep{quinn83, quinn84, DC86, HQ88, HQ89}, although other formation scenarios were proposed (the major merger, \citealp{majorm}, the weak interaction model, \citealp{wim90, wim91}, the space wrapping, \citealp{hq87}). Shells can take on various appearances, but the axially symmetric systems (so-called Type~I shell galaxies, \citealp{wilkinson87}) are particularly easy for studying analytically. We therefore speak only about this type hereafter and assume that they come from a minor merger. They are probably the results of an almost exactly radial collision of a~small galaxy (the secondary) with a~larger and more massive galaxy (the primary). The primary survives the merger almost intact, but the secondary gets tidally disrupted. Shells are density waves made of stars originating in the secondary  that are reaching their apocenters. New shells gradually appear near the center of the primary and expand. Their radii (neglecting their low ellipticity) are determined only by the potential of the host galaxy and the time since releasing the stars from the secondary. A~short recent review of shell galaxies can be found in \citet{bilcjp}.

\begin{figure}
\sidecaption
\resizebox{\hsize}{!}{\includegraphics{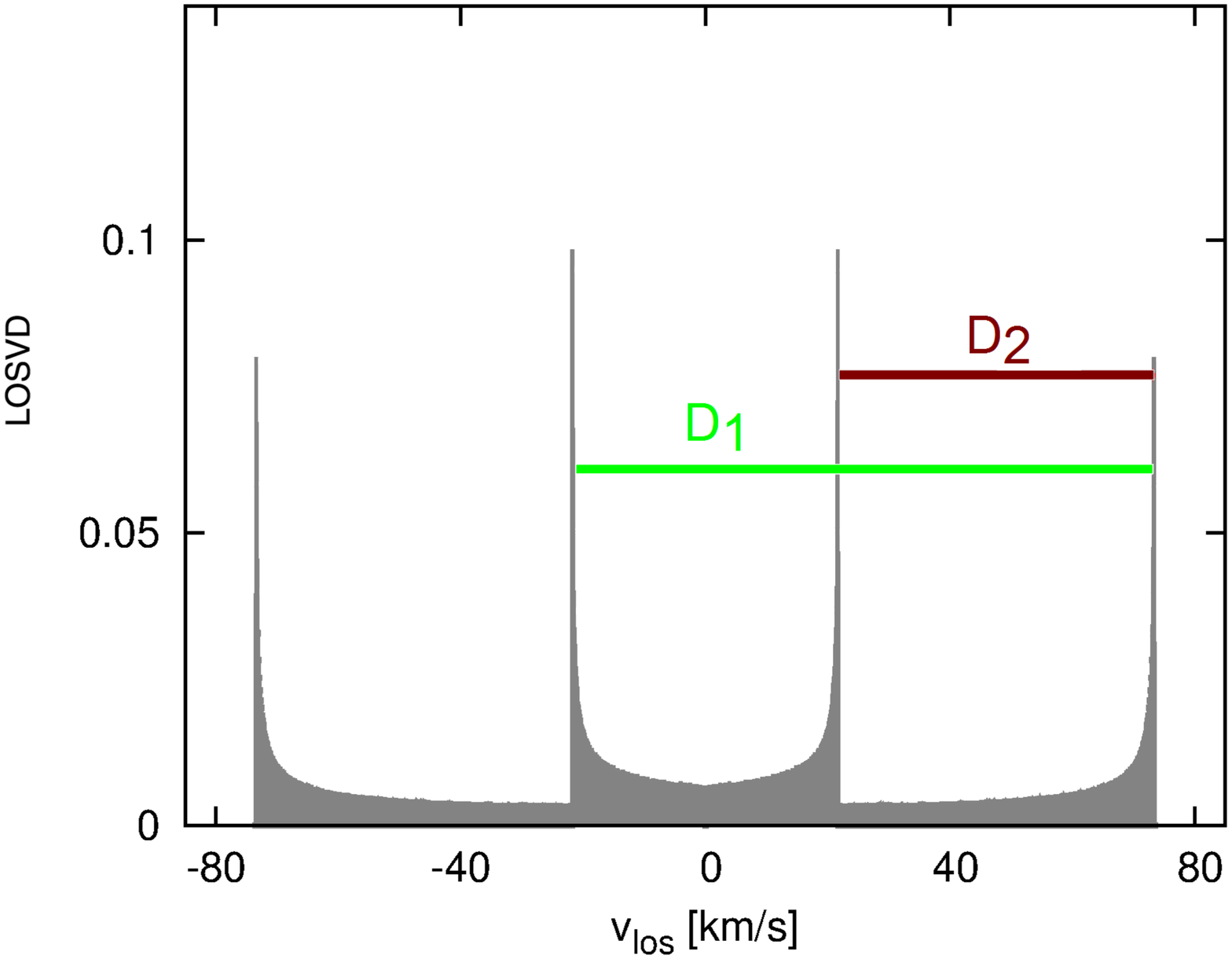}} % one column fig.
\caption{Example of a~shell spectral line profile. The separation $D_1$ is proportional to the circular velocity at the edge of the shell and $D_2$ to the expansion velocity of the shell. See \citet{ebrova12} for details.}
\label{fig:prof}
\end{figure}

Shell kinematics offer a~way to investigate the potential of elliptical galaxies up to large radii \citep{ebrova12, sandhel, fardal}, in some cases up to $\sim$100\,kpc, where the shells occur. \citet{jilkovaprofile}  predict that absorption and emission lines in the spectra of shells have a~quadruple-peaked profile (see \fig{prof}).  The separations between the peaks directly imply the circular velocity at the edge of the shell and the expansion velocity of the shell itself \citep{ebrova12}, see \fig{prof}. However, these observations seem to be just out of reach of current instruments (see chapter 13.4 in \citealp{ebrovadiz}). The main problem is the faintness of shells. The brightest known shells have around 24\,mag\,arcsec$^{-2}$ in the $R$~band. The required spectral resolution is tens of
km\,s$^{-1}$ (see Fig.\,16 in \citealp{ebrova12}). Observing them will
hopefully be feasible with the next 
generation of instruments, such as WEAVE/William Herschel Telescope or OPTIMOS-EVE/E-ELT \citep{wsabstract}.

MOND (MOdified Newtonian Dynamics) was proposed by \citet{milg83a} to solve the mass discrepancy problem without the need for galactic dark matter \citep{milgped, milgcjp, milgmondlaws, famaey12}. It suggests that the laws of dynamics deviate from those deduced from solar-system-conducted experiments for very low accelerations. MOND postulates that 1) an~additional fundamental physical constant $a_0$  exists with the dimension of acceleration, 2)~the departure from Newtonian dynamics occurs for accelerations (any physical quantity with the dimension of m\,s$^{-2}$) that are lower than about $a_0$ (the so-called deep MOND regime), and 3)~in the deep-MOND regime, if a~system of bodies on trajectories $(\mathbf{r}, t)$ conforms to the dynamical equations of a~MOND theory, then the system on trajectories $\lambda (\mathbf{r}, t)$ must also conform to these equations (so-called space-time scaling invariance). The theoretical motivation for such a~modification is unknown. MOND can be  interpreted as a~modification of gravity or as a~modification of the law of inertia. A~recent review of the theoretical aspects of MOND can be found in \citet{milgcjp}. It is often assumed in MOND that the motion of a~test particle in the field of an isolated  spherically symmetric galaxy satisfies the algebraic relation
\begin{equation}
        a_\mathrm{N}=a\mu\left(a/a_0\right).
        \label{eq:algrel}
\end{equation}
 Here $a$ denotes the actual kinematic acceleration of the particle, $a_\mathrm{N}$ the gravitational acceleration calculated in the classical Newtonian way, and $\mu$ the MOND interpolating function. Its precise form is unknown, but it has to obey the limiting constraints
\begin{equation}
        \mu(x) \approx \left\{ \begin{array}{ll}
       x, &  x \ll 1\\
       1, &  x \gg 1.
       \end{array}
   \right.
\end{equation}
In more general situations, we must opt for one of the MOND theories and use its equations \citep{famaey12,milgcjp}.
In some versions of MOND, the acceleration of a~particle even depends also on the motion history of the particle. This is, for example, the case of modified inertia theories \citep{milgmondlaws}.

 In this paper we investigate what MOND predicts about the quantities that shell spectra will allow us to measure: the circular velocity at the edge of the shell and its expansion velocity. Since the dynamics in MOND is determined by the distribution of observable matter (still assuming \equ{algrel}), the separations of the spectral profile peaks are also determined by the distribution of the observable matter. Such behavior is not a~priori required in the dark matter framework.

The paper is organized as follows. In \sect{velgen}, we recall the derivation of the formula for shell expansion velocities. In \sect{asym}, we analyze the asymptotic properties of the two velocities that shell spectral line profiles enable us to measure. More specifically, we focus on the circular velocity in \sect{asymvc} and on the shell expansion velocity in \sect{asymvs}. We study the validity of the results from \sect{asym} for finitely sized shells in \sect{how}, describing our numerical procedure in \sect{model} and applying it to the circular velocity in \sect{howvc} and the shell expansion velocity in \sect{howvs}. If MOND is supposed to be correct, then it is possible to determine the serial numbers of shells from the spectral line profiles. This fact and its implication for recovering the shell system formation history is briefly discussed in \sect{n}. The asymptotic shell and circular velocities are calculated for several observed shell galaxies in \sect{ex} as an example. Finally, we summarize in \sect{sum}. 

%%%%%%%%%%%%%%%%%%%%%%%%%%%%%%%%%%%%%%%%%%%%%%%%%%%%%%%%%%%%%%%%%%%%%%%%%%%%%%%%%%%%%%%%%
%%%%%%%%%%%%%%%%%%%%%%%%%%%%%%%%%%%%%%%%%%%%%%%%%%%%%%%%%%%%%%%%%%%%%%%%%%%%%%%%%%%%%%%%%
\section{Formula for the shell expansion velocity}
\label{sec:velgen}
Here we review the derivation of the formula for calculating the shell expansion velocity in a~general gravitational potential with spherical symmetry. We denote $t$ as the time elapsed since the passage of the secondary though the center of the primary. Since the shells are made of stars near their apocenters, the radius of the $n$-th shell, $r_{n}$, is approximately given as \citep{hq87}
\begin{equation}
        t = \left(n+\Theta\right)P\left[r_n(t)\right], 
        \label{eq:shposth}
\end{equation}
where the shell serial number $n=0,1,2,\ldots$ denotes the number of finished whole oscillations  since the first pericenter of the star. An oscillation is the movement between two subsequent pericenters. The variable $\Theta$, where $0<\Theta<1$, is the phase of the oscillation at $t=0$. The duration of one oscillation for a~star on a~radial orbit with the apocenter at the radius $r$ in a~spherically symmetrical gravitational potential $\phi$ is given by
\begin{equation}
        P(r) = \sqrt{2}\int_0^r\left[\phi(r)-\phi(x)\right]^{-1/2}\mathrm{d}x.
        \label{eq:per}
\end{equation}
If the star was released from the secondary in the primary's center, then its initial phase is $\Theta= 1/2$. In MOND, small secondaries are expected to be disrupted gradually during the initial infall \citep{milgsh, satelliteefe}. The initial phase may differ for individual stars for this reason. The effect of this process on shell positions has never been investigated in detail. However, if the secondary is more cohesive, then the majority of stars are released from the secondary during its passage through the primary's center; i.e., all the stars have approximately $\Theta = 1/2$. This is probably the case for the rich shell systems (see Sect. 6.4 of \citealp{bil14}) possessing more than about seven shells. We  assume $\Theta = 1/2$ hereafter. Then Eq.~\ref{eq:shposth} takes the form
\begin{equation}
        t = \left(n+1/2\right)P\left[r_n(t)\right], 
        \label{eq:shpos}
\end{equation} 
which is twice the free-fall-time from the radius $r$.
This equation is only approximate for the shell's edge position \citep{DC86, ebrova12, bil13}, but it is sufficiently accurate for calculating shell expansion velocities  \citep[see][Tab. 2]{ebrova12}. The expansion velocity of the $n$-th shell at the radius $r$ can be obtained from Eq.~\ref{eq:shpos} by differentiation with respect to $t$ \citep{quinn84}
\begin{equation}
        v_{\mathrm{e},n}(r) = \frac{1}{(n+1/2)\frac{\d P(r)}{\d r}}.
        \label{eq:vsh}
\end{equation}
This is the equation we use for calculating the shell expansion velocities throughout the paper.

%%%%%%%%%%%%%%%%%%%%%%%%%%%%%%%%%%%%%%%%%%%%%%%%%%%%%%%%%%%%%%%%%%%%%%%%%%%%%%%%%%%%%%%%%
%%%%%%%%%%%%%%%%%%%%%%%%%%%%%%%%%%%%%%%%%%%%%%%%%%%%%%%%%%%%%%%%%%%%%%%%%%%%%%%%%%%%%%%%%
\section{Asymptotic properties of the circular and shell expansion velocities in MOND}
\label{sec:asym}
In the following section, we are interested in the MOND implications for the spectral line profiles of the shells with a very large radius. Throughout the paper, we assume that it is possible to construct the gravitational potential $\phi$ in MOND so that bodies move in it by obeying the standard equation of motion $-\nabla \phi = a$. This excludes the nonlocal MOND theories \citep{milgmondlaws}. We adopt the value of the MOND acceleration constant $a_0 = 1.2\times10^{-10}$m\,s$^{-2}$.

%%%%%%%%%%%%%%%%%%%%%%%%%%%%%%%%%%%%%%%%%%%%%%%%%%%%%%%%%%%%%%%%%%%%%%%%%%%%%%%%%%%%%%%%%
%%%%%%%%%%%%%%%%%%%%%%%%%%%%%%%%%%%%%%%%%%%%%%%%%%%%%%%%%%%%%%%%%%%%%%%%%%%%%%%%%%%%%%%%%
\subsection{Shells allow verifying the baryonic Tully-Fisher relation in elliptical galaxies}
\label{sec:asymvc}

In the dark matter framework, the baryonic Tully-Fisher (TF) relation between the asymptotic ($r\rightarrow\infty$) circular velocity and the baryonic mass of the galaxy has still not been fully explained. It does not naturally arise in the cosmological simulations. To match the theory with observations, one must artificially introduce fine tuning between DM and baryonic processes (e.g.,~\citealp{illustris}). Because of the different formation histories of elliptical and disk galaxies \citep{mo}, it would be surprising if disk and elliptical galaxies lay on the same TF relation in a~universe dominated by dark matter. On the other hand, MOND theories firmly predict that the circular velocity of a~test particle asymptotically  reaches a~value given by the baryonic mass of the galaxy, $M$, regardless of its formation history.  The asymptotic circular velocity in MOND reads as \citep{milg83b}
\begin{equation}
        V_{\mathrm{c}} = \sqrt[4]{a_0GM}.
        \label{eq:vca}
\end{equation}

The validity of the baryonic TF~relation expressed by \equ{vca} has been verified many times in disk galaxies \citep{mcgaugh00, mcgaugh11}, but its tests in elliptical galaxies are rare \citep{atlas3d5} owing to the lack of rotating potential tracers. The gravitational potential in ellipticals is most frequently investigated by Jeans analysis. This method does not allow the potential unequivocally to be recovered because of our ignorance of the anisotropy parameter. Shell spectra do not suffer this insufficiency. When they do become available, they will enable us to directly measure the circular velocity and to test the baryonic TF~relation.

%%%%%%%%%%%%%%%%%%%%%%%%%%%%%%%%%%%%%%%%%%%%%%%%%%%%%%%%%%%%%%%%%%%%%%%%%%%%%%%%%%%%%%%%%
%%%%%%%%%%%%%%%%%%%%%%%%%%%%%%%%%%%%%%%%%%%%%%%%%%%%%%%%%%%%%%%%%%%%%%%%%%%%%%%%%%%%%%%%%
\begin{figure}[t]
\sidecaption
\resizebox{\hsize}{!}{\includegraphics{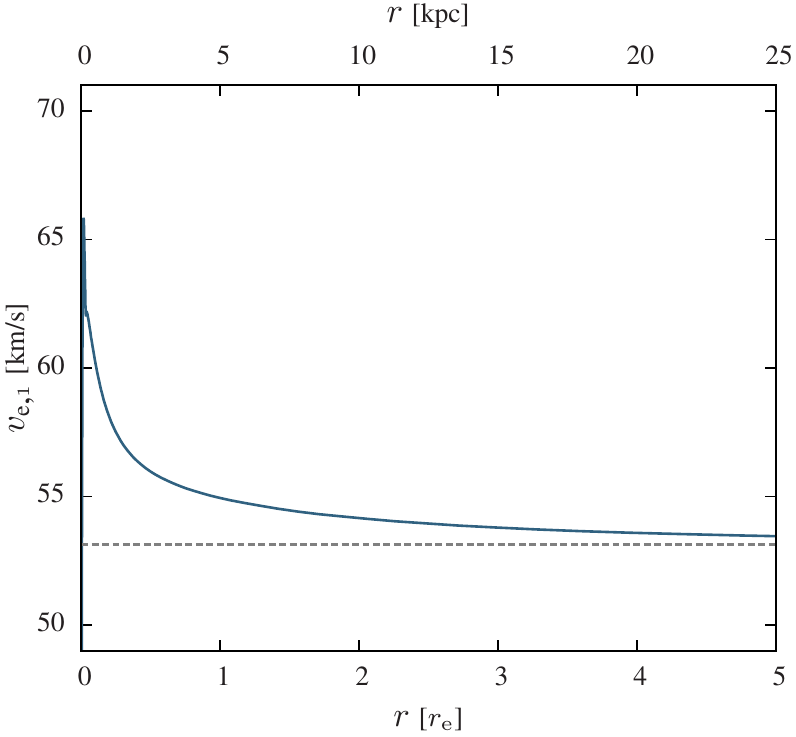}} % one column fig.
\caption{Example of an expansion curve, which is the expansion velocity of a~shell as a function of the galactocentric radius. This curve is drawn for the shell number~1 and the gravitational potential of a de~Vaucouleurs sphere (\sers sphere with index 4). The baryonic mass of the sphere is 10$^{11}$\,M$_\sun$, and it has the effective radius of 5\,kpc. The curve asymptotically reaches the value given by \equ{vsa} indicated by the dashed line.}
\label{fig:expcurve}
\end{figure}

\begin{figure}[b]
\sidecaption
\resizebox{\hsize}{!}{\includegraphics{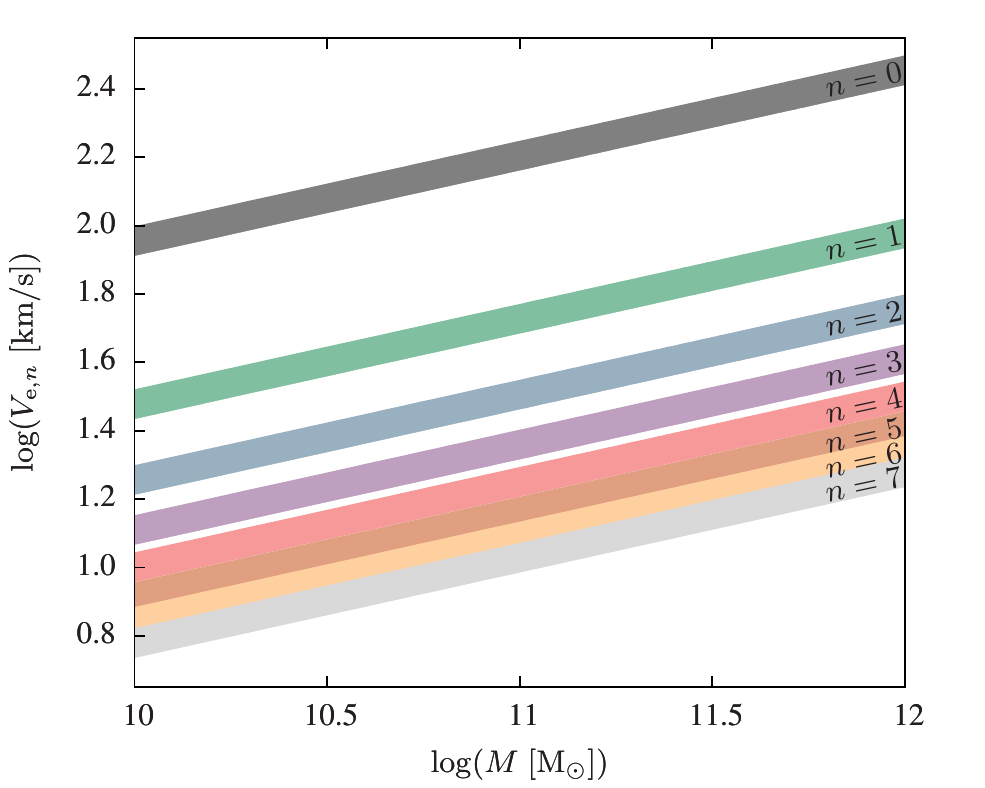}} % one column fig.
\caption{MOND prediction of the relation between shell expansion velocities and their host-galaxy baryonic masses -- a multibranched analogy of the baryonic Tully-Fisher relation. Each branch corresponds to a~certain shell serial number~$n$. The thickness of the drawn branches corresponds to the deviation of $\pm$10\% from the values given by \equ{vsa}.}
\label{fig:tfvs}
\end{figure}

\subsection{Mass vs. shell expansion velocity relation in MOND}
\label{sec:asymvs}
The other quantity that shell spectra allow us to measure is the expansion velocity of the shell. Given a~potential, we can predict the expansion curve (i.e., the shell expansion velocity as a~function of the galactocentric radius) exploiting \equ{vsh}. The expansion velocity of a~shell depends on the whole profile of the gravitational potential from the center of the galaxy up to the radius where the shell edge is located and the serial number of the shell. Similar to in the case of the asymptotic circular velocity, the deep-MOND space-time scaling invariance also implies that the  shell expansion velocity becomes asymptotically constant.  If we investigate the oscillation period at the radius $r$, we can calculate the shell expansion velocity using \equ{vsh}.

The gravitational acceleration in MOND is $a=\sqrt{GMa_0}/r$ for large radii $r$ and the gravitational potential is $\sqrt{GMa_0}\ln{r} + \phi_0$. According to \equ{per} for this logarithmic potential the oscillation period is
\begin{equation}
        P(r) = \frac{\sqrt{2\pi}}{\sqrt[4]{GMa_0}}\,r.
        \label{eq:pr}
\end{equation}
But the real potential of a~galaxy differs from this logarithmic one near the center. However, recall that $P(r)$ is twice the free-fall time.  We can see that a~particle falling from a~very large radius can reach arbitrarily high velocity when it reaches a~given distance from the potential center. Thus, the particle travels across the inner region, where the real potential deviates substantially from the logarithmic one, in an arbitrarily short time. For this reason, the period time for a~real MOND potential is also given by \equ{pr} for very large $r$. Finally, \equ{vsh} indicates that the expansion velocity of the $n$-th shell  asymptotically reaches the value of 
\begin{equation}
        V_{\mathrm{e},n} = \frac{1}{(n+1/2)\frac{\d P}{\d r}(\infty)} = \frac{\sqrt[4]{GMa_0}}{\sqrt{2\pi}\left(n+1/2\right)},
\end{equation}
or, in the terms of the asymptotic circular velocity, 
\begin{equation}
        V_{\mathrm{e},n} = \frac{V_c}{\sqrt{2\pi}\left(n+1/2\right)}.
        \label{eq:vsa}
\end{equation}
\noindent
We show an example of a~shell expansion curve for a~galaxy with the de~Vaucouleur's profile, $M=10^{10}$\,M$_\sun$, and an effective radius of 5\,kpc in \fig{expcurve}. 

\begin{figure}[t]
\sidecaption
\resizebox{\hsize}{!}{\includegraphics{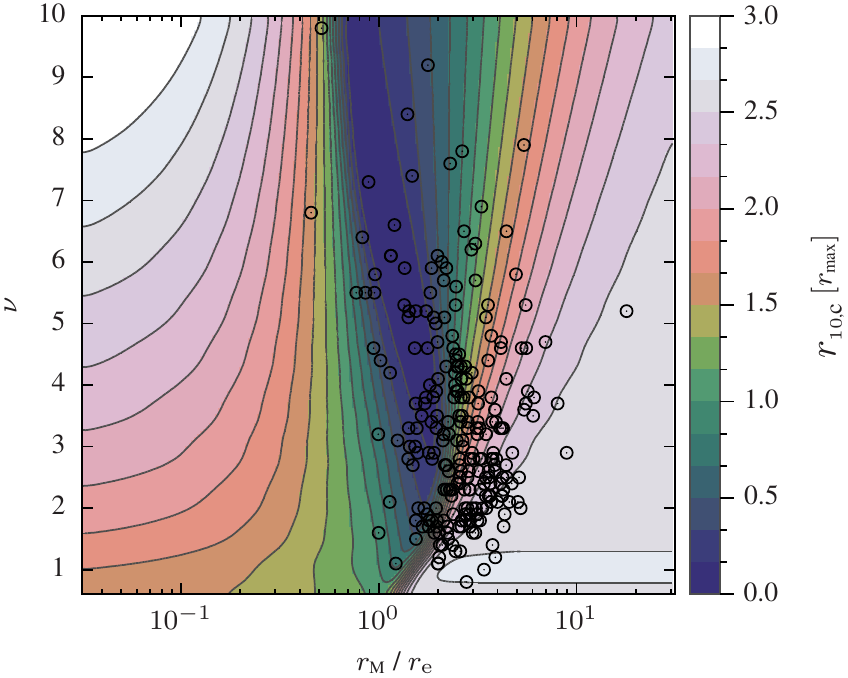}} % one column fig.
\caption{Radius $r_{10,\mathrm{c}}$ expressed in units of $r_\mathrm{max}$ as a~function of the galaxy mass, effective radius $r_\mathrm{e}$, and \sers index $\nu$ (the function $z_{10,\mathrm{c}}(\theta, \nu)$ in \equ{z10}). Above this radius, circular velocity differs by less than 10\% from its asymptotic value. The MOND transitional radius $r_\mathrm{M}$ is connected to the mass of the galaxy by \equ{rm}. The radius $r_\mathrm{max}$ is defined as the maximum of $r_\mathrm{e}$ and $r_\mathrm{M}$. The ratio $r_\mathrm{M}/r_\mathrm{e}$ can be determined observationally from \equ{thobs}. The circles denote the galaxies from the \atlas sample of local elliptical and lenticular galaxies. }
\label{fig:figvc}

We pretend that we have measured the shell expansion velocity for a~lot of shells in many galaxies and that they are large enough (see Sect.~\ref{sec:how}) that their expansion velocities have already reached their asymptotic values. Then \equ{vsa} indicates that shell velocities and galaxy baryonic masses would form
a multibranched analogy of the baryonic TF relation (\fig{tfvs}). Each branch corresponds to a~certain shell serial number~$n$. Or, in another formulation, if we measured the ratio of the circular and expansion velocity for a~lot of shells, then the histogram of this ratio would form a~series of equidistant peaks. The separations between the neighboring peaks would be~$\sqrt{2\pi}$.

 %%%%%%%%%%%%%%%%%%%%%%%%%%%%%%%%%%%%%%%%%%%%%%%%%%%%%%%%%%%%%%%%%%%%%%%%%%%%%%%%%%%%%%%%%
 %%%%%%%%%%%%%%%%%%%%%%%%%%%%%%%%%%%%%%%%%%%%%%%%%%%%%%%%%%%%%%%%%%%%%%%%%%%%%%%%%%%%%%%%%
 
%-----------------------------------------------------------
\end{figure}%---------------------------------------------------------------
\begin{figure}[t!]
\sidecaption
\resizebox{\hsize}{!}{\includegraphics{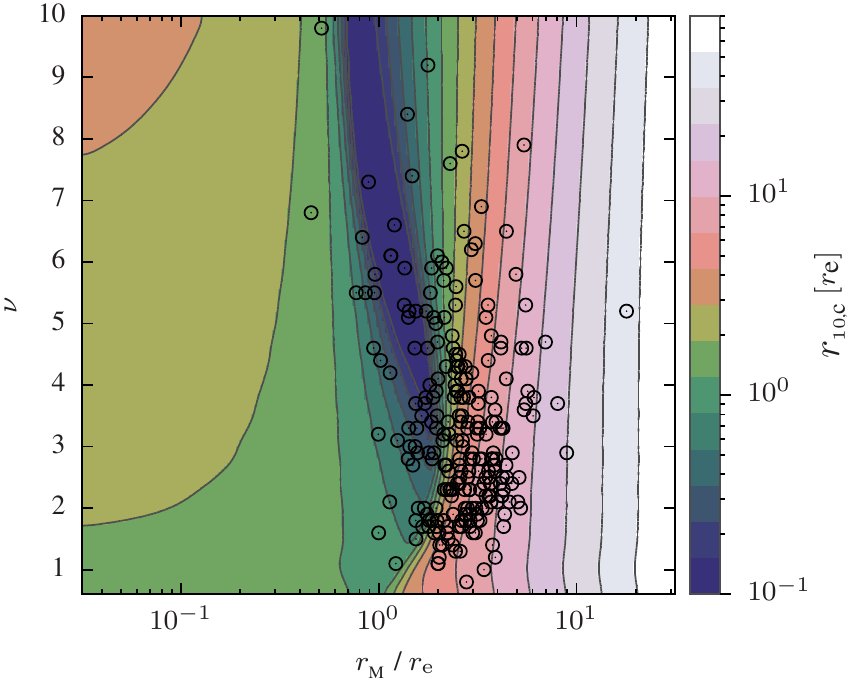}} % one column fig.
\caption{Radius $r_{10,\mathrm{c}}$ in units of the effective radius $r_\mathrm{e}$ (the function $y_{10,\mathrm{c}}(\theta, \nu)$ in \equ{r10}). Similar to \fig{figvc}. Measuring the circular velocity above $r_{10,\mathrm{c}}$ ensures that we can verify observationally the baryonic Tully-Fisher relation without introducing systematic errors that are too large. }
\label{fig:figvcr}
\end{figure}
%---------------------------------------------------------------------------

\section{Where are the asymptotic relations for the circular and shell expansion velocity valid?}
\label{sec:how}
In the previous section, we derived results that are exactly valid only for infinitely large shells. But the real galaxies have only finitely sized shells. Thus we can never measure the asymptotic velocities. This effect brings a~systematic error into the mass-velocity relations found in the previous section. For shell velocities, this effect causes blending of the neighboring branches. To verify the predicted mass-velocity relations observationally, we must use only sufficiently large shells.

In this section, we are interested in the radii $r_{10,\mathrm{c}}$ and $r_{10,\mathrm{e}}$ beyond which the circular velocity and shell expansion velocity, respectively,  differ by less than 10\% from their asymptotic values. The symbol $r_{10}$ stands for $r_{10,\mathrm{c}}$ or $r_{10,\mathrm{e}}$ anywhere we can speak about them  interchangeably. The choice of the deviation lower than 10\% allows us to discern the branches corresponding to the shells numbers 0-3 in the mass vs. shell expansion velocity relation (see \fig{tfvs}) -- supposing that \equ{vsh} works precisely and the observational errors in the galaxy mass and shell velocities are negligible. The radius $r_{10}$ depends, of course, on the distribution of the baryonic matter. We investigated  $r_{10}$ for a~grid of S{\' e}rsic spheres.

\subsection{Modeling the rotation and expansion curves}
\label{sec:model}
The S{\' e}rsic profile describes the surface density of most elliptical galaxies and bulges of disk galaxies well.  The S{\' e}rsic surface density profile  \citep{sersic} is defined as
\begin{equation}
        \Sigma(r) = \Sigma_\mathrm{e}\exp\left\{-b_\nu\left[\left(\frac{r}{r_{\mathrm{e}}}\right)^{1/\nu}-1\right]\right\}.
        \label{eq:sers}
\end{equation}
Here $\nu$ denotes the \sers index: the degree of concentration, $r_{\mathrm{e}}$, the effective radius that encloses one half of the total mass (in projection) and $\Sigma_\mathrm{e}$ the surface density at the effective radius. The term $b_\nu$ is a~function of the \sers index $\nu$. It must be chosen so that the definition of $r_{\mathrm{e}}$ is valid.

\begin{figure}[t]
\sidecaption
\resizebox{\hsize}{!}{\includegraphics{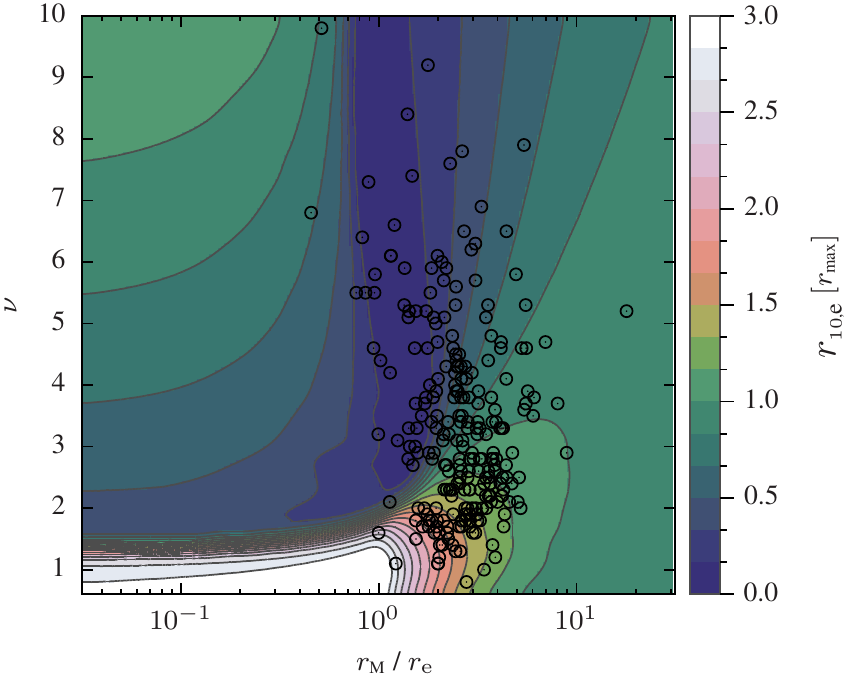}} % one column fig.
\caption{Radius $r_{10,\mathrm{e}}$ expressed in the units of $r_\mathrm{max}$ as a~function of the galaxy mass, effective radius $r_\mathrm{e}$, and \sers index $\nu$ (the function $z_{10,\mathrm{e}}(\theta, \nu)$ in \equ{z10}). Above this radius, shell expansion velocity differs by less than 10\% from its asymptotic value. The MOND transitional radius $r_\mathrm{M}$ is connected with the mass of the galaxy by \equ{rm}. The radius $r_\mathrm{max}$ is defined as the maximum of $r_\mathrm{e}$ and $r_\mathrm{M}$. The ratio $r_\mathrm{M}/r_\mathrm{e}$ can be determined  observationally from \equ{thobs}. The circles denote the galaxies from the \atlas sample of local elliptical and lenticular galaxies.  }
\label{fig:figvs}
\end{figure}

The problem of finding $r_{10}$ must be solved numerically. For this purpose, we created a~grid of \sers spheres. We used the approximate analytic deprojection of the \sers profile proposed by \citet{sersdeproj} with numerical constants  updated by \citet{sersdeprojupdate} 
\begin{equation}
        \rho(r) = \rho_0\left(\frac{r}{c}\right)^{-p}\exp\left[-\left(\frac{r}{c}\right)^{1/\nu}\right], 
        \label{eq:rho}
\end{equation}
where
\begin{equation}
p = 1.0 - 0.6097/\nu + 0.05563/\nu^2,   
\end{equation}
\begin{equation}
\rho_0 = \frac{M}{4\pi\nu c^3\Gamma\left[(3-p)\nu\right]},
\end{equation}
and
\begin{equation}
c = \frac{r_{\mathrm{e}}}{\exp\left[\left(0.6950+\ln\nu\right)\nu-0.1789\right]}.
\end{equation}
The variable $M$ denotes the total baryonic mass of the galaxy.

To recalculate the Newtonian gravitational acceleration to the MONDian one, we used the algebraic relation (\equ{algrel}). We used the interpolating function
\begin{equation}
        \mu(x) = \frac{x}{x+1}.
\end{equation}

The radius $r_{10}$ is clearly a~function  of the parameters of the \sers sphere $r_{\mathrm{e}}$, $\nu$, $M$ and of the basic physical constants $G$ and~$a_0$. From the dimensional grounds, this function  has to have the form of
\begin{equation}
        r_{10} = r_{\mathrm{e}}\,y_{10}\left(\theta, \nu\right),
        \label{eq:r10}
\end{equation}
where
\begin{equation}
        \theta = \frac{r_{\mathrm{M}}}{r_{\mathrm{e}}}
        \label{eq:th}
.\end{equation}
The variable 
\begin{equation}
        r_{\mathrm{M}} = \sqrt{\frac{GM}{a_0}}
        \label{eq:rm}
\end{equation}
is the MOND transitional radius.
After that we reveal the function $y_{10}$, which will allow us to determine practically the radius $r_{10}$ in the units of $r_\mathrm{e}$ once $\theta$ and $\nu$ are measured. This function is discussed in Sects.~\ref{sec:howvc} and~\ref{sec:howvs}.

\begin{figure}[t]
\sidecaption
\resizebox{\hsize}{!}{\includegraphics{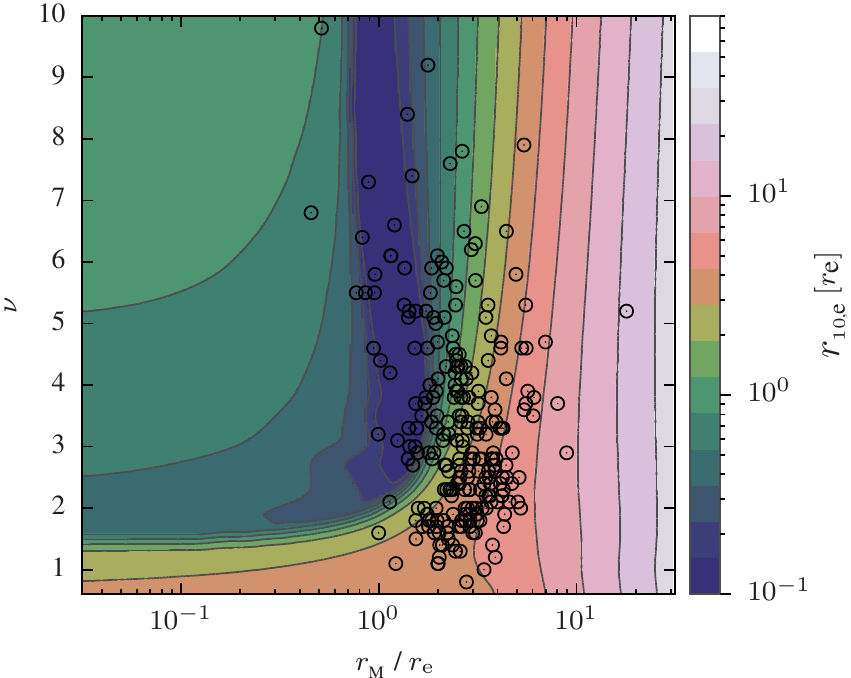}} % one column fig.
\caption{Radius $r_{10,\mathrm{e}}$ in the units of the effective radius $r_\mathrm{e}$ (the function $y_{10,\mathrm{e}}(\theta, \nu)$ in \equ{r10}). Similar to \fig{figvs}. Measuring the circular velocity above $r_{10,\mathrm{e}}$ ensures that we are able to verify the mass vs. shell expansion velocity relation observationally  without introducing systematic errors that are too large. }
\label{fig:figvsr}
\end{figure}

From the theoretical point of view, another function is more interesting than $y_{10}$. One naively expects that $r_{10}$ is greater than both $r_\mathrm{e}$ and $r_\mathrm{M}$. It should be greater than $r_\mathrm{e}$ to enclose the majority of the galaxy's mass and greater than $r_\mathrm{M}$ because the MOND effect that causes the asymptotic flattening gets into action for $r>r_\mathrm{M}$.
Therefore we rearrange \equ{r10} to the form
\begin{equation}
        r_{10} = r_\mathrm{max}\, z_{10}\left(\theta, \nu\right),
        \label{eq:z10}
\end{equation}
where
\begin{equation}
        r_\mathrm{max} = \max\left(r_{\mathrm{e}}, r_{\mathrm{M}}\right).
        \label{eq:rmax}
\end{equation}
The function $z_{10}\left(\theta, \nu\right)$ is discussed in Sects.~\ref{sec:howvc} and~\ref{sec:howvs}.

According to \equ{r10}, we do not have to deal with many combinations of $M$ and $r_{\mathrm{e}}$ when recovering the functions $y_{10}$ and $z_{10}$, but it is enough to vary only $\theta$, which is the ratio of $r_\mathrm{M}$ (determined by $M$) and $r_\mathrm{e}$. The parameter $\theta$ was varied between $10^{-1.5}$ and $10^{1.5}$ and $\nu$ thorough the range in which the approximate deprojection works well, which is 0.6 to 10 \citep{sersdeproj}. We calculated the expansion and rotation curve for each of these \sers spheres and looked for the radius $r_{10}$.

It is well known that the \sers parameters of  galaxies correlate with each other \citep[e.g.,][]{sersphp}. Our grid therefore includes such galaxy profiles that do not actually occur in nature or are very rare. To show which parts of our maps of the functions $y_{10}$~and $z_{10}$~are populated by real galaxies, we also plotted the \atlas galaxy sample in them. This is a~complete sample of 260 well-studied elliptical and lenticular galaxies closer than 42\,Mpc and with stellar mass greater than $6\times 10^9$\,M$_\sun$ \citep{atlas3d1}.  We used the tabled values of their absolute magnitudes in the K band \citep{atlas3d1} to estimate their masses. We assumed the mass-to-light ratio of 1 in the K band for all the galaxies. We used the values of $\nu$ and $r_{\mathrm{e}}$ from the \sers fits by \citet{atlas3d17}.

The parameter $\theta$ can be estimated easily for an observed galaxy. When its image is fitted by the \sers profile, $\theta$ can be calculated as
\begin{equation}
        \theta = \frac{703}{r_{\mathrm{e}}[\arcsec]}\,\Upsilon^{1/2}\,10^{\left(M_\sun-m\right)/5},
        \label{eq:thobs}
\end{equation} 
where $\Upsilon$ is the mass-to-light ratio of the galaxy in the solar units, $M_\sun$ the absolute magnitude of the Sun, and $m$ the apparent magnitude of the galaxy.
 
It is worth noting that $\theta^2$ is proportional to the average surface brightness of the galaxy $L/2\pi r_{\mathrm{e}}^2$. 
%---------------------------------------------------------------------------
%---------------------------------------------------------------------------
\subsection{Circular velocity}
\label{sec:howvc}
Circular velocity measured at the radius $r$ is generally equal to
\begin{equation}
        v_c(r) = \sqrt{r a(r)},
        \label{eq:vcpart}
\end{equation}
where $a(r)$ denotes the acceleration at the radius $r$, supposing isolated spherically symmetric mass distribution. 

The map of the function $z_{10, c}$  ($r_{10,\mathrm{c}}$ expressed in the units of $r_\mathrm{max}$) of the variables $\theta$ and $\nu$ is shown in \fig{figvc}. The radius $r_{10, c}$ is typically between 1~and~2\,$r_\mathrm{max}$ for our models, as well as for the \atlas sample. Interestingly, $r_{10,c}$ is very low or numerically indistinguishable from zero for some of our models. This occurs when the effective radius is comparable with the MOND transitional radius and the \sers index is greater than about~2. In such cases, the galaxy mass profile is in balance with the transition from the Newtonian to the deep-MOND regime. We can see that the galaxies with the rotation curve flat from the center really occur in the \atlas sample.

When testing the baryonic TF~relation in elliptical galaxies (not only using shell spectroscopy), one should use only the measurements of the circular velocity at the radii greater than~$r_{10,\mathrm{c}}$ in order not to introduce too large systematic errors. For the purposes of the practical determination of the radius~$r_{10,\mathrm{c}}$, we generated \fig{figvcr}, where the radius~$r_{10,\mathrm{c}}$ is drawn in the units of~$r_{\mathrm{e}}$. For a~number of \atlas objects, the rotation curve approaches its asymptotic value at many effective radii. This is not very surprising if we note that such object have $\theta\gg 1$; i.e., the deep-MOND regime is reached at many effective radii. These objects correspond to high surface brightness galaxies.

\subsection{Shell expansion velocity}
\label{sec:howvs}
The expansion velocity of the $n$-th shell at the~galactocentric radius~$r$, $v_{s, n}(r)$, can be calculated using \equ{vsh}. The sought function $z_{10,\mathrm{e}}$ ($r_{10,\mathrm{e}}$ expressed in the units of $r_\mathrm{max}$) of the parameters $\theta$ and $\nu$ is shown in \fig{figvs}. We can see that $r_{10,\mathrm{e}}$ is often comparable to~$r_\mathrm{max}$. A~region in this map also exists where $r_{10,e}$ is very low so that the velocity of shells already reach its asymptotic value in the center of the galaxy. This occurs when the effective radius is comparable to the MOND transitional radius and the \sers index is greater than about~2. 

When verifying observationally the mass vs. shell expansion velocity relation, one should use only the shells greater than~$r_{10,\mathrm{e}}$ in order not to introduce systematic errors that are too large (see the introduction of \sect{how}). For the purposes of the practical determination of the radius~$r_{10,\mathrm{e}}$, we generated \fig{figvsr}, where the radius~$r_{10,\mathrm{e}}$ is drawn in units of~$r_{\mathrm{e}}$. For most of our modeled \sers spheres,~$r_{10,\mathrm{e}}$ is lower than~$r_{10,\mathrm{c}}$; i.e., the expansion velocity reaches its asymptotic value typically at a~lower radius than the circular velocity. Again, for the high surface brightness galaxies with high $\theta$, the shell expansion velocity reaches the asymptotic value at more than ten effective radii.

 %%%%%%%%%%%%%%%%%%%%%%%%%%%%%%%%%%%%%%%%%%%%%%%%%%%%%%%%%%%%%%%%%%%%%%%%%%%%%%%%%%%%%%%%%
 %%%%%%%%%%%%%%%%%%%%%%%%%%%%%%%%%%%%%%%%%%%%%%%%%%%%%%%%%%%%%%%%%%%%%%%%%%%%%%%%%%%%%%%%%
\section{Revealing the shell formation history from spectra}
\label{sec:n}
Given an image of a~shell galaxy, what is the serial number of a~particular shell is not evident. If we believe that MOND works in elliptical galaxies, then we are able to derive the gravitational potential from the distribution of the observable matter. This enables us to determine the serial number of a~shell from its spectral line profile. Indeed,  \equ{vsh} indicates that
\begin{equation}
        n = \left[v_{s,n}(r)\frac{\d P(r)}{\d r}\right]^{-1}-\frac{1}{2}.
\end{equation}
If we know the serial number of a~shell, we can derive the time since the decay of the  secondary using \equ{shpos} (or a~more precise one, see \citealp{ebrova12} or \citealp{bil13}). If the ages of all shells in the system are measured, we would be able to reveal the number of shell generations (see, e.g., \citealp{bil13, cooper, DC87, salmon90}) present in the system. Then we could determine the original direction of the arrival of the secondary. The functionality of the  shell identification method \citep{bil13, bil14} could also be independently tested in this way.

 %%%%%%%%%%%%%%%%%%%%%%%%%%%%%%%%%%%%%%%%%%%%%%%%%%%%%%%%%%%%%%%%%%%%%%%%%%%%%%%%%%%%%%%%%
  %%%%%%%%%%%%%%%%%%%%%%%%%%%%%%%%%%%%%%%%%%%%%%%%%%%%%%%%%%%%%%%%%%%%%%%%%%%%%%%%%%%%%%%%%    
\section{Expected asymptotic expansion and circular velocities for several observed shell galaxies}
\label{sec:ex}
In this section, we roughly estimate the asymptotic circular and shell expansion velocities for several real Type-I shell galaxies assuming MOND gravity. The shells in galaxies  NGC\,3923 (e.g., \citealp{prieur88}) and NGC\,7600 (e.g., \citealp{turnbull}) have been known for a~long time. In spite of their prominence, the shells in NGC\,410 and NGC\,3099 have not been reported yet.

For obtaining the surface density profile of NGC\,3923, we fitted the \sers profile to its near-infrared image taken by the Spitzer space telescope. We used the same data, $M/L$ ratio, and the galaxy distance from Earth as in \citet{bil13}.

For NGC\,7600, NGC\,410, and NGC\,3099, we started from their SDSS (Sloan Digital Sky Survey, see, e.g., \citealp{sdssdr10}) images in the $g$, $r$, and $i$ bands. We fitted each of them by the one-component \sers profile in Galfit \citep{Penggalfit}. The integrated magnitudes in the $g$, $r$, and $i$ bands, and the corresponding color indices were determined from these data. The approximate $M/L$ ratios were obtained from the color indices using the tables of \citet{sdssml}. We assumed a~constant $M/L$ across the whole volume of the galaxies.  We get the Galactic extinction coefficients and the distances of these objects from the NASA/IPAC Extragalactic Database. We used the \sers\!-profile parameters $r_{\mathrm{e}}$ and $\nu$ from the fit in the $g$ band. 

We also tried to determine the number of shells greater than $r_{10,\mathrm{e}}$ and  $r_{10,\mathrm{c}}$. For the galaxies NGC\,410 and NGC\,3099, we determined those numbers by visual inspection of their SDSS images. For the galaxies NGC\,3923 and NGC\,7600, we adopt their shell radii lists from \citet{bil13} and \citet{turnbull}, respectively.
The results are shown in Table~\ref{tab:tab}.

\begin{table}[htbp]
\caption{Physical characteristics and predicted circular and expansion velocities for  a~few shell galaxies.}
\begin{tabular}{lllll}\hline\hline
NGC  &410 & 3099 & 3923 & 7600 \\\hline
$D$ [Mpc] & 66.1 & 207 & 23 & 44.2 \\
$r_{\mathrm{e}}$ [kpc] & 11 & 20 & 26 & 4.9 \\
$M$ [10$^{11}$\,M$_\sun$]& 4.9 & 7.8 & 7.8 & 1.2 \\
$\nu$ & 3.8 & 3.9 & 6.1 & 3.7 \\
$r_{\mathrm{M}}$ [kpc] & 24 & 30 & 31 & 12 \\
$\theta$ & 2.3 & 1.5 & 1.2 & 2.4 \\
$V_{\mathrm{c}}$ [km\,s$^{-1}$] & 298 & 334 & 336 & 208 \\
$V_{\mathrm{e}, 0}$ [km\,s$^{-1}$] & 238 & 267 & 268 & 166 \\
$V_{\mathrm{e}, 1}$ [km\,s$^{-1}$] & 79 & 89 & 89 & 55 \\
$V_{\mathrm{e}, 2}$ [km\,s$^{-1}$] & 48 & 53 & 54 & 33 \\
$V_{\mathrm{e}, 3}$ [km\,s$^{-1}$] & 34 & 38 & 38 & 24 \\
$r_{10,\mathrm{c}}$ [$r_\mathrm{e}$] & 2.0 & 0.2 & 0.0 & 2.4 \\
$r_{10,\mathrm{e}}$ [$r_\mathrm{e}$] & 1.2 & 0.0 & 0.0 & 1.4 \\
$N(>\!r_{10,\mathrm{c}})$           & 6   &  2-3 & 27 & 6 \\
$N(>\!r_{10,\mathrm{e}})$           & 8   &   2-3 & 27 & 13 \\\hline
\end{tabular}
\tablefoot{$D$ -- Distance of the galaxy from Earth; $r_\mathrm{e}$ -- Effective radius; $M$ -- Stellar mass of the galaxy; $\nu$ -- \sers index; $r_\mathrm{M}$ -- MOND transitional radius; $\theta$ -- Parameter defined by \equ{th}; $V_{\mathrm{c}}$ -- Asymptotic circular velocity; $V_{\mathrm{e}, n}$ -- Asymptotic expansion velocity of the $n$-th shell; $r_{10,\mathrm{c}}$ -- Minimum large-enough radius for measuring the asymptotic circular velocity; $r_{10,\mathrm{e}}$ -- Minimum large-enough radius for measuring the asymptotic expansion velocity; $N(>\!r_{10,\mathrm{c}})$ -- Number of shells suitable for verifying the baryonic TF relation; $N(>\!r_{10,\mathrm{e}})$ -- Number of shells suitable for verifying the mass vs. shell expansion velocity relation.}
\label{tab:tab}
\end{table}

%%%%%%%%%%%%%%%%%%%%%%%%%%%%%%%%%%%%%%%%%%%%%%%%%%%%%%%%%%%%%%%%%%%%%%%%%%%%%%%%%%%%%%%%%
%%%%%%%%%%%%%%%%%%%%%%%%%%%%%%%%%%%%%%%%%%%%%%%%%%%%%%%%%%%%%%%%%%%%%%%%%%%%%%%%%%%%%%%%%
\section{Summary}
\label{sec:sum}
Spectral lines of stellar shells in symmetrical shell galaxies are predicted to have a~quadruple-peaked profile (see \fig{prof}, \citealp{jilkovaprofile}), and the separations between the peaks can be used for measuring the circular velocity at the edge of the shell and the expansion velocity of the shell itself \citep{ebrova12}. Such a measurement will, however, probably become feasible when  better instruments are available because of the faintness of the shells. In the present paper, we were interested in the MOND predictions regarding these two types of velocities for very large shells. Both these velocities asymptotically reach constant values (\sect{asym}). The asymptotic value of the circular velocity is given by \equ{vca} (known from the early days of MOND) and that of the shell expansion velocity by \equ{vsa}  (new result). 

The formula for the asymptotic circular velocity expresses the baryonic Tully-Fisher relation. MOND predicts that it holds true equally for both disk and elliptical galaxies. In the dark matter framework, nothing like that is  expected. Tests of the baryonic Tully-Fisher relation in elliptical galaxies are rare. When shell spectra become available, they will help us to improve this insufficiency.

From the formula for the asymptotic expansion velocity,
we expect in MOND that the baryonic masses of the shell galaxies, together with the expansion velocities of their shells, form a~multibranched analogy of the baryonic Tully-Fisher relation (\fig{tfvs}). Each branch corresponds to a~certain shell serial number.

We investigated at which galactocentric radius the circular and expansion velocities approximately reach their asymptotic values. Only the shells larger than this radius can be used for verifying the mass-velocity relations we spoke about without introducing large systematic errors. For the circular velocity, this applies for all types of measurement, not only those using the shell spectra. More specifically, we investigated the radius $r_{10}$, above which the circular or expansion velocity differs by less than 10\% from its asymptotic value. The radius $r_{10}$ was sought numerically for a~grid of  \sers spheres (\sect{how}). The map of $r_{10}$ as a~function of the \sers index, and  the galaxy's mass and effective radius is presented as \fig{figvcr} for the circular velocity and as \fig{figvsr} for the shell expansion velocity. Since not all \sers spheres that can be constructed theoretically occur in nature, we also plotted the \atlas sample of local elliptical and lenticular galaxies into these figures. We can see that the rotation or expansion curve is constant with radius for many of these objects, while for others the curves reach their asymptotic values at many effective radii. 

We illustrated our results on the shell galaxies NGC\,410, 3099, 3923, and 7600 (\sect{ex}). For these objects, we roughly estimated  the asymptotic circular and expansion velocities predicted by MOND, the radii $r_{10}$, and the number of shells that are suitable for testing the mass-velocity relations for the circular and shell expansion velocities (\tab{tab}). 

If MOND proves to work in elliptical galaxies, then the shell serial numbers can be derived from the shell spectra, too. The knowledge of the shell serial numbers in a~galaxy allows  many  details about the shell system formation to be revealed including the time since the merger (\sect{n}).

To reach these results, we assumed that the shells stem from a~radial minor merger, that the disruption of the smaller galaxy is most efficient when it goes through the larger galaxy's center, and that the~gravitation potential can be defined in MOND.

\begin{acknowledgements}  
We are grateful for the Czech support for the long-term development of the research institution RVO67985815. MB~acknowledges the project SVV-260089 by the Charles University in Prague.
%%%%%%%%%%%%%%%%%%%%%%%%%%%%%%%%%%%%%%%%%%%%%%%%%%
Funding for SDSS-III has been provided by the Alfred P. Sloan Foundation, the Participating Institutions, the National Science Foundation, and the U.S. Department of Energy Office of Science. The SDSS-III web site is \url{http://www.sdss3.org/}. SDSS-III is managed by the Astrophysical Research Consortium for the Participating Institutions of the SDSS-III Collaboration, including the University of Arizona, the Brazilian Participation Group, Brookhaven National Laboratory, Carnegie Mellon University, University of Florida, the French Participation Group, the German Participation Group, Harvard University, the Instituto de Astrofisica de Canarias, Michigan State/Notre Dame/JINA Participation Group, Johns Hopkins University, Lawrence Berkeley National Laboratory, Max Planck Institute for Astrophysics, Max Planck Institute for Extraterrestrial Physics, New Mexico State University, New York University, Ohio State University, Pennsylvania State University, University of Portsmouth, Princeton University, the Spanish Participation Group, University of Tokyo, University of Utah, Vanderbilt University, University of Virginia, University of Washington, and Yale University. 

\end{acknowledgements}

\bibliographystyle{aa}
\bibliography{citace}

\end{document}